\documentclass[structabstract]{aa}  
\usepackage{graphicx}
\usepackage{natbib}
\usepackage{subfigure}
\usepackage{txfonts}
\begin{document}
   \title{High Resolution CO Observation of Massive Star Forming Regions}

    \author{P. D. Klaassen
          \inst{1}
          \and
          C. D. Wilson\inst{2}
          \and
          E. R. Keto\inst{3}
          \and
          Q. Zhang\inst{3}
          \and
          R. Galv\'an-Madrid\inst{3,4,5}
          \and
          H-Y. B. Liu\inst{3,5,6}
          }

   \institute{European Southern Observatory, Karl-Schwarzschild-Strasse 2,
   		Garching, Germany\\
              \email{pklaasse@eso.org}
         \and
             	Dept of Physics and Astronomy, McMaster University, Hamilton, Canada
          \and
          	Harvard-Smithsonian Center for Astrophysics, Cambridge, USA
	\and
		Centro de Radioastronomia y Astrofisica UNAM, Morelia, Mexico
	\and
		Academia Sinica, Institute of Astronomy \& Astrophysics, Taipei, Taiwan
	\and
		Department of Physics, National Taiwan University, Taipei, Taiwan
             }

   \date{}

% \abstract{}{}{}{}{} 
% 5 {} token are mandatory
 
  \abstract
   { To further understand the processes involved in the formation of massive stars, we have undertaken a study of the gas dynamics surrounding three massive star forming regions.  By observing the large scale structures at high resolution, we are able to determine properties such as driving source, and spatially resolve the bulk dynamical properties of the gas such as infall and outflow. }
  % {Through understanding the effects that massive protostars have on their environments we can place observational constraints on models of massive star formation.}
  {With high resolution observations, we are able to determine which of the cores in a cluster forming massive stars is responsible for the large scale structures. }
   {We present CO observations of three massive star forming regions with known HII regions and show how the CO traces both infall and outflow.  By combining data taken in two SMA configurations with JCMT observations, we are able to see large scale structures at high resolution. }
   {We find large (0.26-0.40 pc), massive (2-3 M$_\odot$) and energetic (13-17 $\times10^{44}$ erg) outflows emanating from the edges of two HII regions suggesting they are being powered by the protostar(s) within.  We find infall signatures in two of our sources with mass infall rates of order 10$^{-4}$ M$_\odot$ yr$^{-1}$.}
   {We suggest that star formation is ongoing in these sources despite the presence of HII regions. We further conclude that the source(s) within a single HII region are responsible for the observed large scale structures; that these large structures are not the net effect of multiple outflows from multiple HII regions and hot cores.}


   \maketitle
%
%________________________________________________________________

\section{Introduction}
\label{sec:intro}

The formation mechanism responsible for low mass stars like the Sun is, in broad terms, well constrained, from pre-stellar cores \citep{Ward02} to pre-main-sequence stars \citep[i.e.][]{Shu87,Andre2000}.  At approximately 8 M$_\odot$ however, the situation becomes more  complicated.  The forming star begins to burn hydrogen hot and fast enough for ultraviolet radiation to push outwards and ionize the surrounding gas \citep[although this may be shifted to higher starting masses by puffing up of the protostar, see][]{Hosokawa09,Yorke08}.  With these extreme outward pressures, how then can accretion onto the forming star continue beyond the formation of an HII region?  Does it halt before the HII region expands to observable sizes \citep[i.e.][]{GL99,msc}? Does it occur in self shielding toroids around the massive protostar \citep[i.e.][]{Keto07,Peters10,liu10_toroid}? Or, does a molecular accretion flow transition into an ionized one at the boundary of the HII region \citep[i.e.][]{Keto02,Keto03}?  If accretion (either molecular or ionized) continues beyond the formation of an HII region, there should be observational evidence for this. It could come in the form of ionized/molecular infall, or large scale outflows that are still being powered.

Outflows act as an important release mechanism for the buildup of angular momentum in star forming regions on all mass scales \citep{Richer99,Arce07}. They can also be some of the largest scale phenomena associated with star formation, and are therefore easier to observe at the large distances to massive star forming regions. The outflows observed in nearby low mass star forming regions can be quite collimated, while collimation factors in higher mass star forming regions do not appear to be as high.  This last point has  led some authors to suggest that as intermediate and high mass stars evolve up the main sequence \citep[i.e.][]{KW06}, the outflows lose collimation because they are older and the outflow has had more of a chance to widen into the surrounding medium \citep[i.e.][]{Beuther05}.  Indeed, there are no observations to date of highly collimated outflows for sources with luminosities greater than 10$^5$ L$_\odot$ \citep[masses greater than O7 main sequence stars;][]{Mckee07}.  It should be noted that the highest mass stars are intrinsically further away, but that the loss of collimation with increasing source luminosity appears not to be due to beam dilution \citep[i.e.][]{Beuther05}.  Through characterizing the kinematics of the gas surrounding the formation sites of massive stars, we can begin to constrain the formation mechanisms which can produce these large scale phenomena (i.e. how much gas is accelerated into the outflow).  Once gas kinematics are determined for a number of sources, we can begin to constrain models of massive star formation (i.e. give realistic parameter space boundaries for phenomena on observable scales).

The large average distances to massive star forming regions and the clustered and embedded nature of massive star formation requires that these regions are observed with interferometers at sub-millimeter and longer wavelengths in order to spatially resolve the kinematics of the gas in which massive protostars are embedded.  With interferometers, we can resolve the spatial structures and kinematics of a number of molecular species and often determine whether the spectral line signatures are coming from the massive protostar/HII region or neighboring hot cores. In general, interferometers filter out emission from large scale structures and can be quite useful for observing small scale structures within larger scale emission \citep[i.e.][]{Eisner08}.  With the Submillimeter Array\footnote{The Submillimeter Array is a joint project between the Smithsonian Astrophysical Observatory and the Academia Sinica Institute of Astronomy and Astrophysics and is funded by the Smithsonian Institution and the Academia Sinica.}  (SMA) in its extended configuration, at 230 GHz, when structures are smaller than $\sim10''$, there is no need to include large scale structure information from a more compact configuration or single dish telescope since most of the flux is recovered by the interferometer.

In the general interstellar medium, $^{12}$CO is very abundant and can be excited into some of its lowest rotational energy states (i.e. J=1 or J=2) at the fairly low temperatures that  can be reached in molecular clouds. Bulk motions of the gas, such as the material entrained into an outflow from a forming star, can thus be  traced on large scales by CO emission.  With our combined interferometer and single dish datasets, we can observe these large scale structures at high resolution.

Here, we present 1mm observations of $^{12}$CO and $^{13}$CO towards three massive star forming regions; G10.6-0.4, G28.20-0.05, and NGC 7538 IRS1 (hereafter G10.6, G28.2, and NGC 7538).  Our observations were obtained with the SMA in its extended configuration in late 2005, with compact configuration observations taken in 2007 and 2008.  Single dish $^{12}$CO observations were obtained at the James Clerk Maxwell Telescope (JCMT)\footnote{The James Clerk Maxwell Telescope is operated by The Joint Astronomy Centre on behalf of the Science and Technology Facilities Council of the United Kingdom, the Netherlands Organisation for Scientific Research, and the National Research Council of Canada.} in 2006 (project M06BC09).  In Section \ref{sec:observations} we present our observations and image combination method, while we present the results of our kinematic studies in Section \ref{sec:results}.  In Sections \ref{sec:discussion} and \ref{sec:conclusions} we discuss our results and conclude. 

\begin{table*}
\caption{Observed Sample of Massive Star Forming Regions.}
\label{tab:sources}
\centering
\begin{tabular}{lllccccc}
\hline\hline
Source &\multicolumn{2}{c}{pointing center (J2000)} &\multicolumn{2}{c}{V$_{\rm LSR}$}&\multicolumn{2}{c}{Distance}\\
\cline{2-3} \cline{4-5} \cline{6-7}\\
& RA & DEC & (km s$^{-1}$) &ref & (kpc) & ref\\
\hline
G10.6-0.4	&18:10:27.77 & $-$19:56:04.5& -3 & 1& 6.0	&4	\\
G28.20-0.05	&18:42:58.17 & $-$04:13:57.0& 99 &2& 5.7&2 	\\
NGC7538 IRS1&23:13:45.30 & $+$61:28:10.0& -59 &3& 2.7	&5	\\
\hline
\end{tabular}
%\tablefoot{}
\tablebib{(1)~\citet{Purcell06}; (2) \citet{Pestalozzi05}; (3) \citet{Zheng01}; (4) \citet{KZK08}; (5) \citet{Moscadelli09}}
\end{table*}

\section{Observations}
\label{sec:observations}

Because CO is one of the most abundant molecules in the interstellar medium (ISM), its emission is generally quite extended and much of the flux can be filtered out by an interferometer.  Thus, we observed it in two SMA configurations (extended and compact) to observe structures on size scales ranging from $\sim 1''$ to $\sim 25''$.  CO emission may be more extended than this, and to ensure that the large scale outflows are properly characterized, we obtained maps of these regions at the JCMT.  

The SMA extended configuration data were obtained in September 2005 with CO (J=2-1, 230.538 GHz) in the upper sideband at a uniform spectral resolution of $\sim$ 0.53 km s$^{-1}$.  The extended configuration data was reduced using the SMA version of MIRIAD \citep[as was done in][]{K09}.

\begin{table*}
\caption{SMA and JCMT Observation Characteristics}
\label{tab:obs}
\centering
\begin{tabular}{lrrrrrrccc}
\hline\hline
Source &	\multicolumn{2}{c}{ Gain Calibrators} & Res.\tablefootmark{a}	& \multicolumn{3}{c}{RMS noise (Jy/beam)} & \multicolumn{3}{c}{ Synthesized Beam\tablefootmark{b}}	\\
\cline{2-3}\cline{5-7}\cline{8-10}	
& extended & compact & (km s$^{-1}$) & SMA\tablefootmark{c} & JCMT\tablefootmark{d} & Comb.\tablefootmark{e} & Major\tablefootmark{b} & Minor\tablefootmark{b} & PA\tablefootmark{b}\\
\hline
 G10.6	&	1833-210	&	1733-130	&	0.63	&						0.24	&	0.9	&	0.24	&	1.41$''$&1.14$''$ &54.4$^\circ$	\\
G28.2	&	1944-201	&	1733-130	&	0.63	&						0.12	&	1.2	&	0.14	&	1.88$''$ & 1.07$''$& 77.9$^\circ$	\\
NGC 7538	&	2202+422	&	0102+584\tablefootmark{f}	&	1.06	&	0.11&	2.0	&	0.13	& 	2.35$''$ & 1.61$''$ &71.6$^\circ$	\\
\hline
\end{tabular}
\tablefoot{}
\tablefoottext{a}{Spectral resolution of the final data products.}
\tablefoottext{b}{Synthesized beam of the combined SMA and JCMT dataset. }
\tablefoottext{c}{Jointly inverted SMA extended and compact configuration data.}
\tablefoottext{d}{JCMT observations regridded to the spatial resolution of the jointly inverted SMA observations.}
\tablefoottext{e}{RMS noise value for the combination of the JCMT data and the jointly inverted SMA data using the miriad task MOSMEM at the spectral resolution of the final data product.}
\tablefoottext{f}{BLLac was also used as a gain calibrator, and 3c273 and Uranus were both used as bandpass calibrators in addition to 3c454.}
\end{table*}

Our SMA compact configuration data was obtained in September 2007, June 2008, and July 2007 (respectively for the source order in Table \ref{tab:sources}).  The spectral resolution was the same for G10.6 and G28.2 (0.53 km s$^{-1}$), but only half for NGC 7538 (1.06 km s$^{-1}$).  Just like the extended configuration data, bandpass calibration was done using 3c454.5, while gain calibration was done using the sources listed in Table \ref{tab:obs}.  The compact configuration data were reduced using the SMA native MIR IDL package, exported to MIRIAD and self-calibrated before being combined with the extended configuration data. For NGC 7538, a two point mosaic was created in the compact configuration instead of a single pointing as was done for the other sources. Their pointing centers  are RA (2000) = 23$^{\rm h}$13$^{\rm m}$43$^{\rm s}$.75, DEC (2000) = $+$61$^{\circ}$28$'$21$''$.49 and RA (2000) = 23$^{\rm h}$13$^{\rm m}$46$^{\rm s}$.75, DEC (2000) = 61$^{\circ}$27$'$59$''$.49.

JCMT observations were obtained for $^{12}$CO in October 2006 using RxA.  7$\times$7 pixel raster maps were made at 10$''$ spacings at a spectral resolution of 0.63 km s$^{-1}$ over the full 1 GHz JCMT bandwidth. Each map took one hour to complete and the rms noise levels for each map are listed in Table \ref{tab:obs} in units of Jy beam$^{-1}$ regridded to the resolution of the combined SMA datasets.  At the native resolution of the JCMT observations the rms sensitivity limits were 0.09 K in the T$_{\rm A}^*$ scale at a spectral resolution of 0.63 km s$^{-1}$. The main beam efficiency for these observations was 0.69. For each source, multiple JCMT raster maps were coadded and linear baselines removed from each pixel.  Data reduction was completed using the JCMT starlink software and the datacubes were exported to MIRIAD for combination with the SMA observations.

\subsection{Combining Datasets}
\label{subsec:combining}

During the formation of stars of all masses, the angular momentum releasing outflows can become quite extended.  In the study of \citet{Wu04}, only two sources (one classified as high mass, and one classified as low mass) have angular sizes of less than 6$''$ (see their Figure 3), and at most, six of the 397 sources in their study would have observable in the $\lesssim10''$ scales to which our highest resolution SMA observations are sensitive.  Thus, to properly understand outflow properties at high resolution, we need to combine high and low resolution datasets to ensure we are probing the gas at all relevant spatial scales. Comparing the integrated flux levels in our CO observations from the SMA to those from the JCMT suggest that our high resolution ($\sim1"$) extended configuration observations recover $\sim$ 5\% of the total flux and our lower resolution ($\sim 4"$) compact configuration data recover $\sim$ 40\% of the flux.  When combined, the SMA datasets appear to recover approximately 50\% of the single dish flux.  

Linear baselines were removed from the JCMT raster maps as part of the data reduction process. Thus, the continuum emission must also be removed from the SMA datasets before combination with the single dish data. This was done in the UV plane.  The two SMA datasets (for a given source) were then simultaneously brought into the image plane using the INVERT command and specifying a spectral resolution of 0.63 km s$^{-1}$ to match the spectral resolution of the JCMT observations.  For NGC 7538, the SMA compact configuration data had a spectral resolution of 1.06 km s$^{-1}$, so, when inverting, this lower spectral resolution was used instead. 

Since combining the JCMT data with that from the SMA is a form of cleaning, both a JCMT image and beam map are required.  The JCMT beam map used for deconvolution was created using the MIRIAD command IMGEN assuming a gaussian source of intensity 1 and a radius of 20.8$''$ (the resolution of the JCMT at 230.5 GHz).  The image cube was then regridded to the resolution of the jointly inverted interferometer image.   Once the single dish and interferometer maps were on the same flux scale and at the same resolution, they were combined using a maximum entropy method (using the MIRIAD command MOSMEM) and a clean image was restored using the interferometer map as the dirty image and the results from MOSMEM for the clean components.
 
For each source, we also tried combining the single dish and interferometer maps by feathering (using the MIRIAD command IMMERGE) which did not recover all of the single dish flux.  We also tried inverting the JCMT observations to the UV plane and jointly inverting all three datasets; however this method only produced reliable results for two out of the three sources, and for consistency, we present our results of the MOSMEM combined datasets for all three sources.    To test the robustness of our image combination, we determined outflow properties from the single dish only data, the single dish and lowest resolution interferometer map, or all three datasets combined.  Using the same spatial and velocity limits, the properties are the same in all three instances.  The values are lower however when we only combine the two interferometer maps because we are filtering out the large scale structures consistent with the flux filtering described earlier in this section. We have also compared our JCMT observations of NGC 7538 to those of \citet{Davis98}. Using their velocity integration limits (-94 to -64 km s$^{-1}$ and -50 to -20 km s$^{-1}$), we compared the fluxes in our map to theirs.  Our map is much smaller than theirs, and they show CO contours well beyond our map boundary. Their emission covers a much greater area than we have sampled with the SMA, and thus we can really only compare peak fluxes between the two maps. We find that our fluxes are somewhat lower than theirs, however both the red and blue peaks are within 15\% of theirs. 

\section{Results}
\label{sec:results}

Here, we present our observational results.  Of the three observed sources, we find indications of large scale molecular outflows in two sources (NGC 7538 and G28.2), as well as molecular infall signatures in two sources (NGC 7538 and G10.6).  We note that while there is no large scale molecular outflow signature coming from G10.6, an ionized outflow has been previously detected in this region \citep{KW06} and despite our non-detection of infall in G28.2, an infall signature has been seen in the dense gas \citep[NH$_3$,][]{Sollins05} surrounding this region.

\subsection{Outflow of molecular gas from the HII regions}
\label{sec:outflow}

\begin{figure*}
\subfigure[G28.2-0.04]{\includegraphics[width=0.5\textwidth]{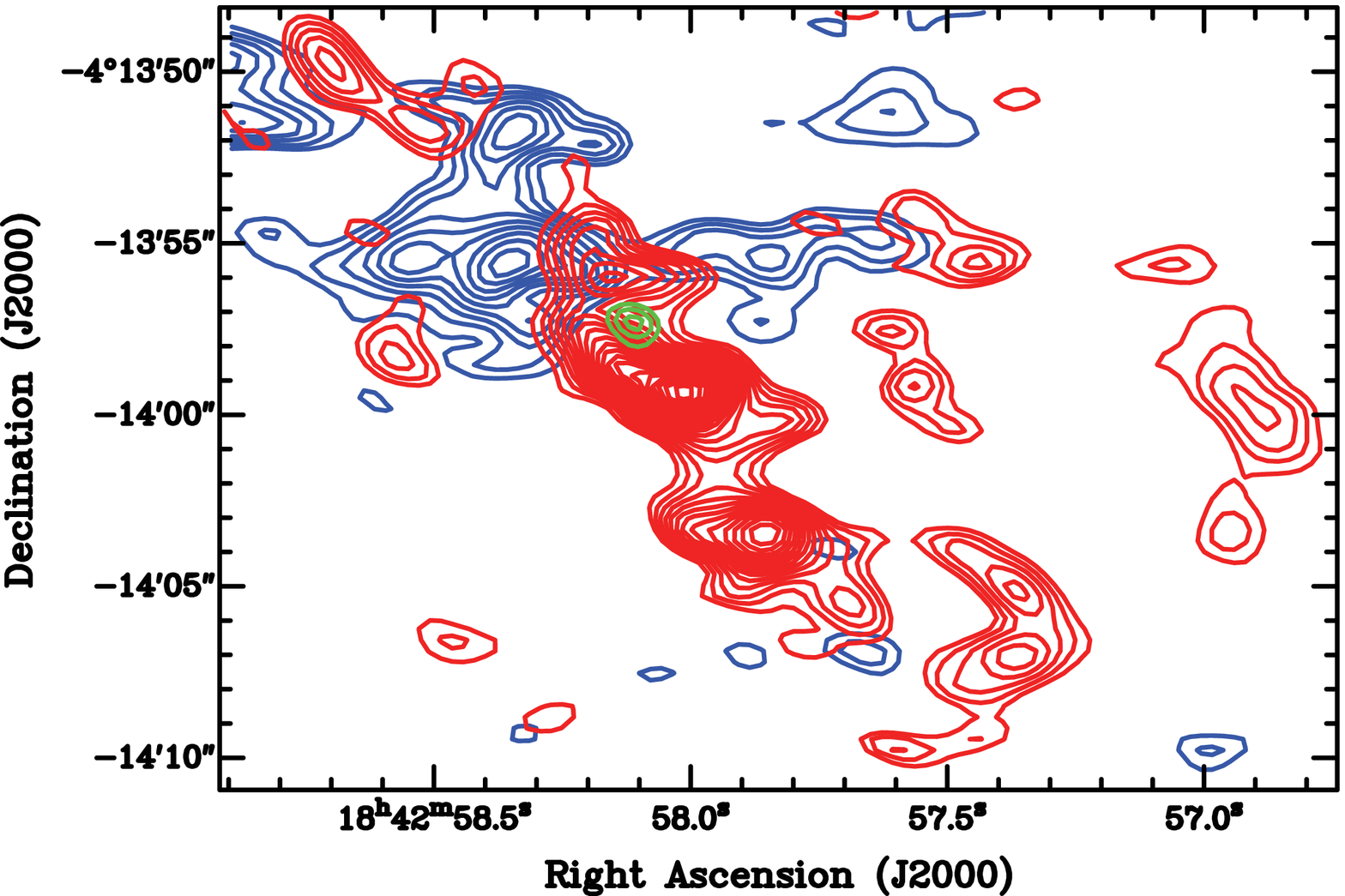}}
\subfigure[NGC 7538 IRS 1]{\includegraphics[width=0.43\textwidth]{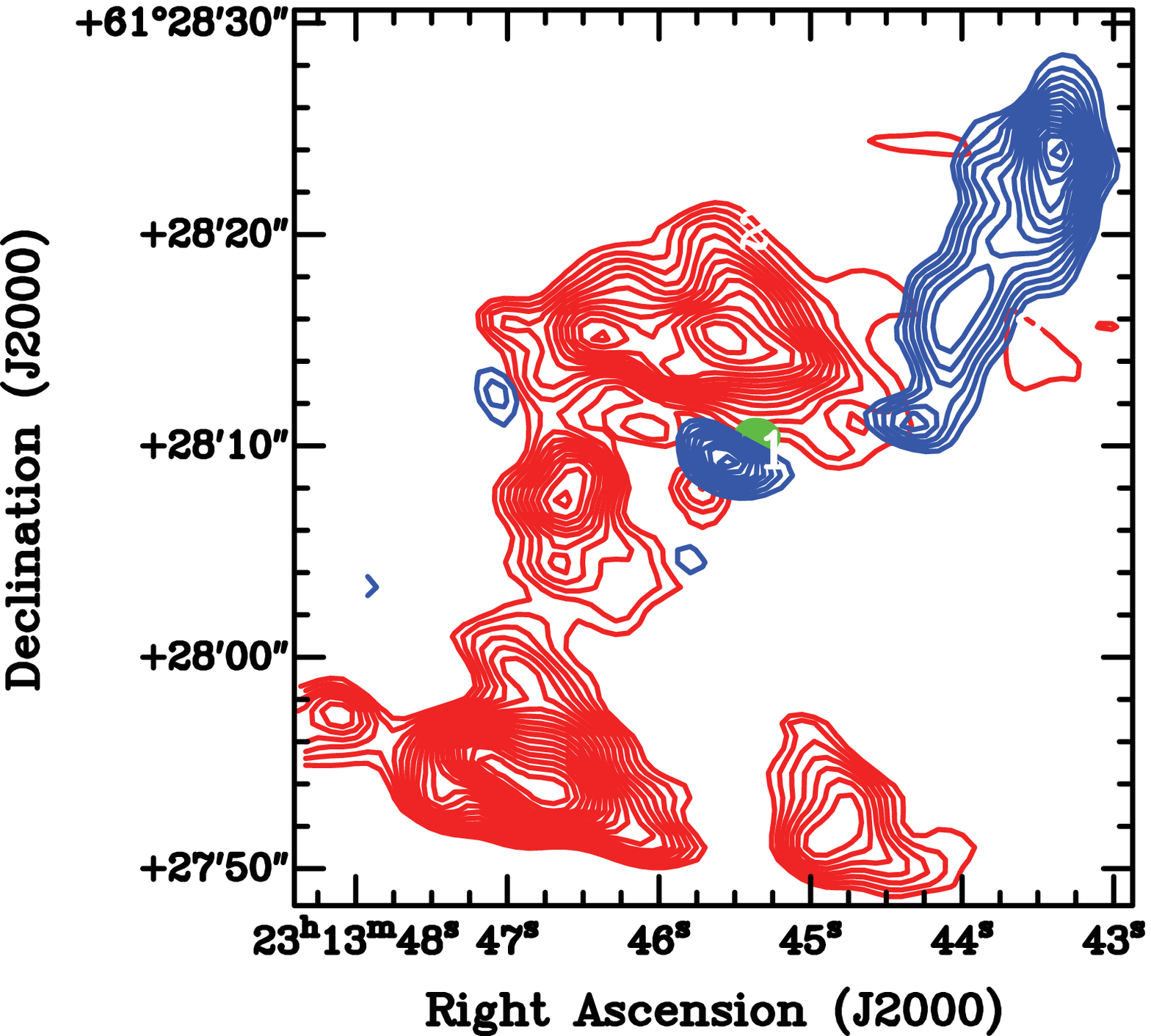}}
\caption{Large-scale Molecular Outflows: In both figures, the blue and red contours show the blue and red shifted CO emission integrated over the velocity ranges given in Table \ref{tab:obs} starting at 5$\sigma$ (3$\sigma$), and increasing in intervals of 1$\sigma$.  The green contours represents the continuum emission, starting at 50\% of the peak emission.    In the right panel, the numbers indicate the locations of IRS 1, 2, and 3.}
\label{fig:outflows}
\end{figure*}

While outflows are generally characterized as one of the larger phenomena associated with star formation, it is only by observing them at high resolution that we can constrain which source is producing the observed large scale outflow.  In each of the cases presented below, when detected, the outflow appears to emanate from a single HII region.   The molecular outflow originates at the boundary of the HII region, suggesting that perhaps the outflow generation is occurring within the HII region.

The three datasets for CO were combined as described in Section \ref{subsec:combining}, and from these datacubes we created zeroth moment (integrated intensity) and first moment (intensity weighted velocity) maps of the red and blue shifted emission independently for each source to determine the outflow dynamics. These maps were created using both velocity range and noise clipping masks.

The high abundance of CO does make it an excellent tracer of high velocity and bulk gas motions; however, it also causes the line to become optically thick quite quickly.  For our calculations we have only integrated over the high velocity gas regions. If we were to integrate over the lower velocity gas, we begin detecting ambient and envelope gas.  Our inner velocity limits were chosen though trial and error.  We started with our low velocity limits quite close to the V$_{\rm LSR}$, and continued moving that limit to higher blue and red shifted velocities until the maps showed primarily outflow (i.e collimated) structures. with little ambient contamination.  Our inner velocity limits are at relatively high velocities, which means we are only integrating over the `outer wings' in the terms of \citep{Margulis85}.  This is a limitation of our observations, and thus we are underestimating the outflow properties.  We have assumed optically thin emission in the wings.

In all of our calculations below, we assume an ambient temperature of 100K. This value was chosen to represent the bulk of the outflowing gas since it is colder than the hot core gas surrounding each of these HII regions \citep[ $>$ 300 K as traced by CH$_3$CN,][]{K09} and warmer than the temperatures found in cold dense cores \citep[$<$ 20 K i.e.][]{Garay04,Peretto10}.  We note that changing the temperature by a factor of two changes our estimates by less than a factor of two.

Integrated intensity maps were created for the sources with outflows and are presented in Figure \ref{fig:outflows}, while the velocity ranges used to create the maps (centered on the local standard of rest velocity (V$_{\rm LSR}$) given in Table \ref{tab:sources}) are presented in Table \ref{tab:results}.  To create the first moment maps and determine the average red and blue shifted velocities, we shifted the velocity axis of each datacube so that the V$_{\rm LSR}$ is centered at 0 km s$^{-1}$, and set range masks over the same velocity ranges used for the zeroth moment maps and clipped the images. 

Using the spatially averaged integrated intensity in both the red and blue outflow lobes, we calculated the average H$_2$ column density of the emitting region by using a constant of proportionality between the column density and integrated intensity (N$=1.3\times 10^{15}\int$Tdv) which assumes a temperature of 100 K.  This average column density, when multiplied by the emitting area and mass of H$_2$ gives the mass in each outflow lobe ($M=NA$) when the abundance of CO with respect to H$_2$ is taken into account \citep[10$^{-4}$,][]{Herbst09}.  Since a first moment map is in essence a measure of velocity, the zeroth and first moment maps were multiplied together to determine the momentum ($P=Mv$) in each outflow lobe.  Similarly, we also determined the kinetic energy in each lobe ($E=1/2*Mv^2=P^2/2M$).  These mass, momentum and energy values are presented in Table \ref{tab:results}.  The quoted uncertainties come from the uncertainties in the integrated intensities and velocities, and are propagated through the equations described above.

\begin{table}
\caption{Outflow Results from Kinematic Analysis}
\label{tab:results}
\centering
\begin{tabular}{lcc|cc}
\hline\hline
&\multicolumn{2}{c}{G28.2-0.04} &\multicolumn{2}{|c}{NGC 7538 IRS 1}\\
&\multicolumn{2}{c}{($t_{\rm dyn}=2.4\times10^4$ yr)} &\multicolumn{2}{|c}{($t_{\rm dyn}=4.2\times10^4$ yr)}\\
&\multicolumn{1}{c}{Blue}&\multicolumn{1}{c|}{Red}&\multicolumn{1}{c}{Blue}&\multicolumn{1}{c}{Red}\\
&\multicolumn{1}{c}{($\Delta v$=-10,-3)}&\multicolumn{1}{c|}{($\Delta v$=3,10)}&\multicolumn{1}{c}{($\Delta v$=-20,-10)}&\multicolumn{1}{c}{($\Delta v$=10,20)}\\
\hline
M\tablefootmark{a}	&	0.89	&	1.60	&		0.43		&	0.98		\\
P\tablefootmark{b}	&	8.3	&	8.7	&		4.6		&	10.7		\\
E\tablefootmark{c}	&	7.6	&	4.9	&		4.9		&	11.8		\\
L\tablefootmark{d}	&	0.30	&	0.20	&		0.10		&	0.23		\\
$\dot{\rm M}$\tablefootmark{e}	&	3.6		&	6.4	&	1.0	&	2.3		\\
\hline
\end{tabular}
\tablefoot{A calibration uncertainty of 20\% is not propagated through these equations.  The velocity ranges listed above each column ($\Delta v$) represent the integration limits with respect to the V$_{\rm LSR}$ stated in Table \ref{tab:sources}.}
\tablefoottext{a}{Mass in outflow lobe in units of M$_\odot$}
\tablefoottext{b}{Momentum in outflow lobe in units of M$_\odot$ km s$^{-1}$}
\tablefoottext{c}{Energy in outflow lobe in units of 10$^{44}$ erg}
\tablefoottext{d}{Mechanical Luminosity in outflow lobe in units of L$_\odot$}
\tablefoottext{e}{Mass outflow rate in units of 10$^{-5}$ M$_\odot$ yr$^{-1}$}
\end{table}

We estimated a kinematic age for G28.2 (t$_{\rm dyn} = 2.4\times10^4$ yr) by dividing the size of the outflow by the velocity of the gas. For NGC 7538, we took a value from the literature \citep[t$_{\rm dyn} = 4.2\times10^4$ yr,][]{Davis98} since larger scale observations of this region show outflow motions well beyond the edges of our primary beam.  Using these kinematic ages, and the derived outflow masses and energies, we determined average mass loss rates ($\dot{M}=M/t$) and outflow mechanical luminosities ($L=E/t$).  These values can also be found in Table \ref{tab:results}.  For both outflow sources, the mass outflow rates are greater than the values generally seen in lower mass star forming regions \citep[which can be in the range 10$^{-8}-10^{-5}$ M$_\odot$ yr$^{-1}$,][]{Shepherd03}.  Assuming the amount of mass in the outflow scales proportionally to the amount of mass infalling onto the star \citep{Beuther02}, then we would expect larger mass outflow rates than for lower mass stars since the masses of the outflows are larger.

For each source, the orientation of the outflows  are generally consistent with being perpendicular to the rotation directions of the hot cores surrounding the HII regions \citep[as traced by OCS and SO$_2$ in][]{K09}, and the outflow kinematics are consistent with other outflows listed in \citet{Wu04} and towards the values they show for high mass star forming regions (those shown with open squares in their figures).  We note that one of our sources overlaps with their study (NGC 7538), and that our derived outflow properties are consistently lower than their values. As discussed in Section \ref{sec:observations}, our map does not cover the entire flow in NGC 7538, and we are thus missing much of the extended flux.  Since mass scales directly with integrated intensity under the assumption of optically thin gas, we expect our outflow masses to be lower than previously derived values \citep[i.e.][who used larger JCMT maps to determine the outflow properties]{Davis98,Qiu11}.  It is likely that this is the case for G28.2 as well.  

The lower derived outflow properties can also be understood in terms of the high optical depths of CO at low velocities.  \citet{Margulis85} discusses how the bulk of the outflow mass and dynamics can be found in the lower velocity wing emission, from what they called the `inner wing'.  \citet{Choi93RMxA} described this same region as containing the `high velocity' gas.  To use the terminology from these papers, we are only tracing the gas in the `outer wing' or `extremely high velocity' regions (respectively).  As stated previously, $^{12}$CO becomes optically thick quickly in massive star forming regions, and we cannot trace the bulk of the gas that is optically thick in this tracer.  The high opacity of the line is advantageous for seeing the extremely high velocity gas, but means that we are missing most of the mass in the outflow.  We cannot comment on the mass of the gas at lower velocities than the limits quoted in Table \ref{tab:results}, but must assume that it is quite large if there is still more than a solar mass of material in the outer wings of these outflows.

\subsection{Infall of Molecular Gas onto the HII regions}

\begin{figure*}
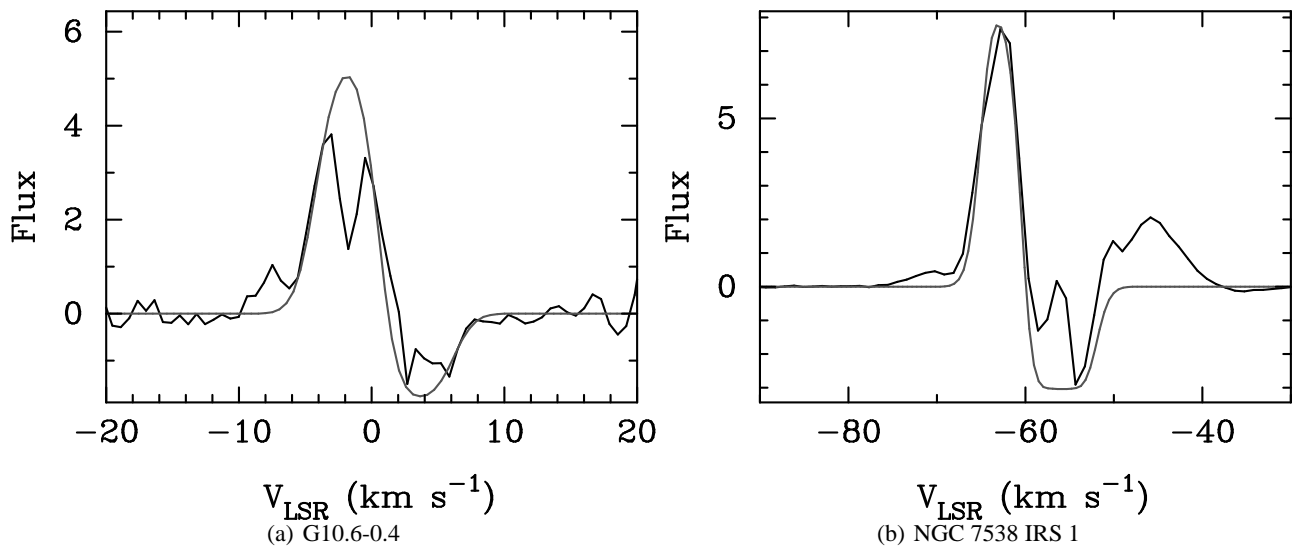

\subfigure[G10.6-0.4]{\includegraphics[width=0.37\textwidth,angle=-90]{g10.pcyg.ps}}
\subfigure[NGC 7538 IRS 1]{\includegraphics[width=0.37\textwidth,angle=-90]{n7.pcyg.ps}}
\caption{CO (J=2-1) spectra integrated over the infall regions overlapping with the HII regions for G10.6 (left) and NGC 7538 (right). The grey lines shows the 2 layer infall model characterized by the parameters given in Table \ref{tab:infall}.}
\label{fig:pcygni}
\end{figure*}

Bulk infall motions can manifest as inverse P-Cygni profiles with redshifted absorption and blueshifted emission \citep[i.e.][]{DiFrancesco01} along the line of sight to the infalling region against a bright background.   From this profile, a rough estimate of the infall velocities and masses can be determined.  \citet{Beltran11} found inverse P-Cygni profiles in $^{13}$CO towards two of the three sources they studied; the third, G10.6, had $^{13}$CO emission on scales much larger than they were able to probe with their data.  For the sources where they do detect infall, they find infall velocities of 3.8 and 5 km s$^{-1}$ for G19.61-0.23 and G29.96-0.02, respectively.  Similarly, inverse P-Cygni profiles have also been observed towards G24.78+0.08 \citep[ 2 km s$^{-1}$ in NH$_3$][]{Beltran06}, G31.41+0.31 \citep[ 3.1km s$^{-1}$ in C$^{34}$S][]{Girart09}, and in W51 IRS2 \citep[ 4 km s$^{-1}$ in CN][]{Zapata08}.   In order to determine whether this infall signature is present towards our sample of sources, we used our combined JCMT and SMA CO data. We note that the infall signatures, when present, appear virtually the same whether or not the JCMT data is included since these signatures occur on small scales (a few arcseconds). We find infall signatures in the CO from G10.6 and NGC 7538, but not towards G28.2.  Below, we describe how these signatures were characterized to determine source properties.

We found the pixel with the largest CO absorption feature along the line of sight to the HII region, and created an integrated intensity map based on the width of the absorption feature.  We determined the  -3$\sigma$ contour of the feature in the moment map, and created an integrated spectrum based on the overlap between that contour, and the surface of the HII region.  For G10.6, the smaller region was the absorption feature, for NGC 7538 it was the unresolved HII region.

Using the two layer infall model of \citet{DiFrancesco01} we determined infall velocities from the inverse P-Cygni profiles in CO for G10.6 and NGC 7538.  The inputs to these two layer models consist of  Planck functions characterized by their different temperatures.  These temperatures correspond to the cold layer of gas in front of the emission region (T$_f$=7.4 K, the excitation temperature of CO J=2-1), to the rear of the emission region (T$_r$) and the temperature of the continuum source (T$_c$). The infall velocities are presented in Table \ref{tab:infall}.  With these infall velocities, known source distances and integration areas, we also determined the mass infall rates using Equation 3 from \citet{KW07}:

\begin{equation}
\dot{M}_{in} = (4/3)\pi n_{H_2}\mu m_H r^2 v_{in}
\end{equation}

where $n_{H_2}$ is the ambient density, set here to 10$^6$ cm$^{-3}$, $\mu$ is the mean molecular weight (2.35), $m_H$ is the mass of a hydrogen atom, $r$ is the radius of the emitting area, and $v_{in}$ is the infall velocity. We suggest that these inverse P-cygni profiles are due to local infall and not foreground absorption because these same structures have been seen in other molecules towards these sources as well.  See for instance \citet{Qiu11} for NGC 7538 and for G10.6 see \citet{Ho86,Keto88}

 For both sources, the mass infall rate was 2$\times10^{-4}$ $M_\odot$ yr$^{-1}$.  For G10.6, the rest velocity required to align the model with the observations is shifted with respect to the local standard of rest velocity by $+$ 4 km s$^{-1}$.  The profiles plotted in Figure \ref{fig:pcygni} only reflect the CO along the line of sight to the HII regions, and not the areas shown in Figure \ref{fig:outflows}.

In both sources, inverse P-Cygni profiles can also be seen in CH$_3$CN (J=12-11) originally presented in \citet{K09}. The derived temperatures (323 and 310 K respectively) and spatial structures of the CH$_3$CN regions in G10.6 and NGC 7538 as described in \citeauthor{K09} suggest that this species is tracing the hot cores surrounding these two HII regions. The profile is strongest in the K=3 transition for both sources, and these infall signatures are consistent with those seen in CO; the 4 km s$^{-1}$ velocity shift in the G10.6 profile included.  That the CH$_3$CN also shows an infall signature suggests that material is still actively accreting onto the two HII regions.  The CH$_3$CN spectra shown in the top and middle panels of Figure 7 of \citet{K09} are integrated over the entire CH$_3$CN emitting region for G10.6 and NGC 7538 respectively, and not just the area showing an inverse P-Cygni profile, yet the signature is still visible in the integrated spectrum from NGC 7538 \citep[see][]{Qiu11}.   The simulations of high mass star forming regions of \citet{Wang10} which include both outflow feedback and magnetic fields find similar mass infall rates, and, as discussed later, this molecular mass infall rate in G10.6 may be comparable to the ionized mass infall rate as traced by H66$\alpha$.

\begin{table}
\caption{Infall Model Parameters from Inverse P-Cygni Profiles}
\label{tab:infall}
\centering
\begin{tabular}{lrrrrr}
\hline\hline
Source Name & T$_{\rm r}$ & T$_{\rm c}$ & V$_{\rm in}$ & V$_{\rm LSR }$ & $\dot{M}_{\rm in}$\\
&(K)& (K) & (km s$^{-1}$) & (km s$^{-1}$)& (M$_\odot$ yr$^{-1}$)\\
\hline
G10.6 & 250 & 170 & 2.5 & 1 &2$\times10^{-4}$\\
NGC 7538 & 385 & 400 &2.3 & -59 & 2$\times10^{-4}$\\
\hline
\end{tabular}
\end{table}

\subsection{Results for Individual Sources}

\subsubsection{NGC 7538 IRS1}

As shown in the right panel of Figure \ref{fig:outflows} both the red and blue outflow lobes appear emanate from  the central HII region (IRS 1), suggesting that the outflow may be precessing or bending around other structures (such as IRS 2).  On much larger scales than observable here, the outflow appears to be dominantly in the North-South direction (S. Corder, priv. comm).  The outflow direction we find is consistent with the previous observations of \citet{Davis98} which were on similar size scales to those observed here, but at much lower resolution.

The energetics from this source are comparable to those in G28.2, despite having a lower bolometric luminosity than G28.2. This is interesting because it is much closer than either G28.2 or G10.6. It is unclear how distance affects our results since the outflow from NGC 7538 would still be resolved at any of the distances to the sources studied here.  The outflow from this source is the  most  extended in this study, with each lobe spanning more than 0.4 pc \citep[at an assumed distance of 2.7 kpc][]{Moscadelli09}.

Comparing our CO map to Spitzer IRAC images of the region, we see that the blue outflow lobe curves around a 3.6 and 5 $\mu$m feature coming from IRS 2 which is not very prominent in the 4.5 $\mu$m band.  This could mean that the CO is tracing the edge of a region that is deficient in molecular gas, and dominated by PAH emission since H$_2$ and CO are expected to emit in the 4.5 $\mu$m band while PAH emission is expected to be more prominent in the 3 and 5 $\mu$m bands \citep[see Figure 1 of ][]{Reach06}.  The portion of the Spitzer map corresponding to the red emission lobe is saturated in all four IRAC bands.   The ionized emission from this region \citep[as traced by H30$\alpha$ see][]{KZK08} is unresolved in these observations and does not show any systematic or bulk motions, so a similar analysis cannot be conducted.

%With a source luminosity of 9.6$\times10^4$ L$_\odot$ \citep{Kraus06}, NGC 7538 IRS 1 is our best example of a large scale, fairly well collimated outflow. The IRS 2 and IRS 3 clumps are not detected in the continuum observations presented here, but are both within our field of view as indicated in Figure \ref{fig:outflows}.

Because NGC 7538 is relatively nearby for a high mass star forming region \citep[2.65 kpc][]{Moscadelli09}, a number of distinct sites of massive star formation have been identified (i.e. IRS 1, 2, and 3).  Both IRS 2 and 3 are within our field of view, as indicated in Figure \ref{fig:outflows}, yet it is clear that the detected large scale outflow is centered on IRS 1.

As stated earlier, the outflow properties derived here are much smaller than those determined in \citet{Davis98}. Judging from Figure 12 of \citeauthor{Davis98}, their intensities were calculated over an area greater than $80"\times80"$, while our integration area was closer to $40"\times40"$ since we only had a single pointing with the SMA in its extended configuration. \citet{Kameya89} observed CO J=1-0 in this source with the NRO 45 m antenna.  Their map of the gas coming from NGC 7538 IRS1 is only slightly more extended than ours, and their derived outflow properties are closer to our values than those of \citeauthor{Davis98}. The maps and derived outflow properties for this source from \citet{Qiu11} are also larger than ours. This suggests that their larger maps are able to detect more of the flux from the outflow. That the outflow properties in \citet{Kameya89} and \citet{Qiu11} are  lower than those of \citet{Davis98} suggests they have also missed part of the outflow. It should be noted that in \citet{Qiu11}, the outflow parameters are calculated over the multiple outflows they suggest for this region.

The continuum brightness for this source at 230 GHz is $\sim$ 2.83 Jy/beam, and the absorption feature is strongest ( -2.9 Jy/beam) at -54 km s$^{-1}$ suggesting that there is complete absorption along the line of sight.  We suggest there is complete absorption over a velocity range of a few km s$^{-1}$ (as shown by the model spectrum) with contaminating emission from the outflow causing the line intensity to increase to near zero at $-56$ km s$^{-1}$.

\subsubsection{G28.20-0.05}

 The 5.8 $\mu$m Spitzer IRAC image of this region shows a slight extension in this region (centered on the HII region) towards the south east of the map in the left panel of Figure \ref{fig:outflows}.   It should be noted that this extension is perpendicular to the outflow shown here and along the rotation direction shown in \citet{K09}, but is much more extended than the warm rotating gas seen in OCS and SO$_2$.  We also note that the radio recombination line \citep[H53$\alpha$,][]{KZK08} data for this source does not appear to have an infall signature, but there does appear to be a velocity gradient across the source consistent with the rotation seen in \citet{K09}.

As was the case for NGC 7538, the outflow moment mapping appears to fit best when the LSR velocity is taken from the average of the values listed in the literature over the last ten years (99 km s$^{-1}$) instead of the one assumed in \citet{K09}.

The outflow kinematics from this source are all slightly lower than the trend shown in \citet{LS09}, suggesting we are perhaps missing some of the gas at high optical depths.  The red and blue outflow lobes span 0.55 and 0.41 pc on the sky (respectively, at an assumed distance of 5.7 kpc) which makes the physical scale of this outflow comparable to NGC 7538.

Curiously, we do not detect an infall signature towards this HII region like we do in the others, despite the infall signature having been seen in NH$_3$ by \citet{Sollins05_g28}.  This may suggest the large scale infall has stopped but that infall is still ongoing in the densest gas surrounding the HII region.

\subsubsection{G10.6-0.4}

An ionized outflow has been seen towards G10.6 in H66$\alpha$ \citep{KW06}, but this is the first high resolution search in this source for a molecular outflow which includes the large scale structures observable with a single dish telescopes.  That there is no evidence for a large scale molecular outflow is highly curious \citep[although see][for a possible explanation]{Liu10,Liu11}.  Previous studies of this region have suggested an orientation that is primarily in the plane of the sky. 

Inverse P-Cygni profiles exist in both CO and CH$_3$CN for this source.  As shown in Table \ref{tab:infall}, the two layer infall model is only able to reproduce the CO inverse P-Cygni profile if the rest velocity is set to 1 km s$^{-1}$ (not -3 km s$^{-1}$ as listed in Table \ref{tab:sources}).  That these two species show this absorption feature slightly redshifted from the source velocity suggests that perhaps the molecular gas is doing something different than the ionized gas. It should be noted that the NH$_3$ observations of \citep{Sollins05} for this source also show a similar velocity shift at high resolution for the molecular gas. The first moment map they present in Figure 2 shows gas between -3 and 4 km s$^{-1}$ suggesting an average velocity of 1 km s$^{-1}$ as well.  \citet{Keto88} also place the central velocity of their NH$_3$ observations at 1 km s$^{-1}$ (see their Figure 1).

There is the posibility that our infall signature is contaminated by absorption from the foreground spiral arm in the approximate velocity range of -10 km s$^{-1}$ to 50 km s$^{-1}$ \citep[see Figure 1 of ][]{Fish03}.  However, that absorption feature is quite flat in HI and thus we suggest it only minimally contributes to our absorption feature.  Our feature has a depth consistent with total absorption over a much smaller velocity range than the 60 km s$^{-1}$ range of the HI absorption suggesting that this is a separate feature  \citep[although see Appendix A of][]{Liu10}. None of the parameters going into our infall model are greatly affected by this absorption as the depth of our absorption feature shows complete absorption.

The molecular infall appears to be consistent with the infall of the ionized gas as traced by H66$\alpha$ \citep{KW06}.  From modeling the infall signature in the ionized gas, \citet{KW06} determined a mass infall rate of 1$\times10^{-3}$M$_\odot$ yr$^{-1}$.  From our two layer infall model, we determined a molecular gas infall velocity, which we input into Equation 3 of \citet{KW07} and determined the molecular gas infall rate to be 2$\times10^{-4}$M$_\odot$ yr$^{-1}$, the same as that inferred from the  NH$_3$ observations of \citet{Ho86}. \citet{Keto90} show a mass infall rate of 2.5$\times10^{-4}$ at a distance of 0.1 pc from the HII region as well.

For comparison, molecular infall has also been detected towards G24.78+0.08 \citep{Beltran06}, and that HII region appears to be undergoing a contraction consistent with an accretion flow onto the HII region as well \citep{GalvanMadrid08}. This suggests that G10.6 is not the only source with these properties, although G24.78+0.08 has been shown to be powering a large scale outflow \citep{Furuya02}.

\section{Discussion}
\label{sec:discussion}

 Here, we only see one outflow emanating from each massive star forming region.  This is not always the case however; in G5.89-0.39, the large scale outflow \citep{msc} is in a different direction than the smaller scale outflows of \citet{Sollins04} and \citet{Hunter08} who also see multiple outflows in various directions \citep[see][for an alternate interpretation of NGC 7538]{Qiu11}.

Figure 5 of \citet{LS09} shows outflow mass, momentum and energy as a function of bolometric luminosity for a large sample of outflows from massive star forming regions observed with the IRAM 30 m telescope.  They find that each of these quantities increases as a function of source bolometric luminosity.  We find that our derived values are lower than their trend, which we suggest is due to using $^{12}$CO instead of $^{13}$CO for our analysis.  We are primarily tracing `outer wing' gas \citep[in the terminology of ][]{Margulis85}, where, by definition, the $^{13}$CO is no longer emitting.  Were we able to distinguish between the ambient and outflow gases at lower velocities, our fluxes would increase, and thus our derived parameters would be much more consistent with those of \citet{LS09}.  An overestimate in the flux will lead to a proportional overestimate in the outflow properties (see the equations given in Section \ref{sec:outflow}).

{The advantage of the higher resolution data is that we are able to determine which source is powering the large scale outflow with much higher precision.  \citet{KK08} at the same resolution as the data presented here, showed that the small scale $^{12}$CO surrounding W51e2 appears at the edges of the HII region (see their Figure 2). The outflow in this region can be traced all the way down to the ionization boundary, despite the presence of multiple HII regions within 10$''$ of each other \citep[see, for instance][]{Zhang97}.}Thus, in order to reveal which source is responsible for the large scale energetics, high resolution observations are crucial.  Here, we have shown that the large scale outflows are produced within a single HII region, and are not the product of the energetics of the surrounding hot cores.  \citet{Peters10_witness} simulated accretion and feedback in a massive star forming region with multiple sink particles representing multiple massive stars. They showed that the single large scale outflow produced from these clusters comes from the most massive member of the cluster.  Our observations seem to be showing the same thing; the large scale outflow coming from the brightest HII region within our field of view.

%None of the sources discussed here are associated with Herbig-Haro objects, suggesting they are all deeply embedded, and not heating up their environments enough to excite H$_2$.  This further suggests that most of the energy is being released in the outflow.  Since we are capturing the outflow down to small size scales, we suggest we are capturing the bulk of the outflow energetics.

  In both cases, we have been able to trace the outflow back to its powering source (the HII region).  That the outflowing gas can be traced back to the edges of a single HII region suggests that the massive protostar(s) within that HII region are powering the outflow.  We also suggest that something inside the HII region is powering the large scale structure because of the continued infall seen towards these sources as well.  This leads us to conclude that the massive, large scale outflows detected in these regions are powered by the massive protostar(s), and are not the averaged gas motions from the entire cluster expelling gas to release angular momentum.  Our ratio of $\dot{M}_{\rm out}$ to $\dot{M}_{\rm in}$  is approximately 0.3, which appears to be consistent with the ratio of the jet entrained outflow mass to mass accretion rate discussed in \citet{Beuther02}, noting that they found a lot of scatter in their value (0.17 for a source whose luminosity is accretion driven).

There are infall signatures for each source, either presented here or in \citet{Sollins05_g28} suggesting continued accretion beyond the formation of the HII region.  In none of our sources is this as clear as in G10.6 where both molecular and ionized infall are detected.  The infalling molecular gas appears to pass through the ionization boundary, become ionized and continue infalling with similar mass infall rates. Spatially, these two infalling populations appear to be nested within each other, supporting this theory further.  These observations further suggest that high mass star formation may be able to proceed in a manner similar to, but scaled up from, lower mass star formation with the added complications of HII regions and clustering.

It is interesting that we do not detect an infall signature in G28.2, where one has previously been detected in the dense gas \citep[in NH$_3$, see][]{Sollins05_g28}.  This suggests that perhaps the large scale collapse has finished, but infall and accretion are ongoing in the dense gas close to the HII region which we cannot trace with CO or CH$_3$CN.    Oddly, this source is our youngest source (based on a kinematic age for the outflow) as well as the one with the most collimated outflow. Further investigation will be necessary to understand why infall was not detected in CO.

\section{Conclusions}
\label{sec:conclusions}

We combined observations of CO (J=2-1) from two SMA configurations (extended and compact) with observations of the large scale CO structures observed at the JCMT for three massive star forming regions.  We detected large scale molecular outflow signatures from two of them as well as molecular infall signatures towards two sources one of which has previous detections of an ionized infall in H66$\alpha$.  

Due to the high resolution of our datasets we have shown that the outflowing gas originates at the edges of the HII regions. This suggest  that the outflows are driven from a single powering source and are therefore not the spatially averaged outflows from a star forming cluster.   Our data also suggest that these outflows are continuing to be powered despite the presence of the HII regions.

If these outflows are powered by accretion, which is likely the case, the continued presence of powered outflows is an indication that accretion is ongoing as well, despite the presence of an HII region. This is further supported by the detection of infall signatures in the molecular gas towards two of our sources which small scale infall has been detected in denser tracers by other authors for the third source. From inverse P-Cygni profiles, we infer infall velocities and masses much higher than for lower mass star forming regions, but consistent with those found in regions forming high mass stars.  We also note that, for G10.6, the molecular and ionized infall rates for the ionized and molecular gas even appear to be similar.

\begin{acknowledgements}

The authors would like to thank the referee and the editor for their comments which helped improve the paper.  

\end{acknowledgements}

\bibliography{bibtest}{}
\bibliographystyle{aa}

\end{document}